\documentclass[sigconf]{acmart}
\usepackage{booktabs} % For formal tables

\usepackage{tabularx}
\usepackage{caption}
\usepackage{algorithm}
\usepackage{varwidth}
\usepackage{algpseudocode}
\usepackage{multirow}
\newcolumntype{I}{!{\vrule width 3pt}}
\newlength\savedwidth

\newlength\savewidth

\graphicspath{{./}}

\begin{document}

\copyrightyear{2019}
\acmYear{2019}
\setcopyright{acmcopyright}
\acmConference[SEC 2019]{The Fourth ACM/IEEE Symposium on Edge
Computing}{November 7--9, 2019}{Arlington, VA, USA}
\acmDOI{10.1145/3318216.3363373} 
\acmISBN{978-1-4503-6733-2/19/11}
\title[\small{Task-Adaptive Incremental Learning for Intelligent Edge Devices}]{Task-Adaptive Incremental Learning for Intelligent Edge Devices}
\author{Zhuwei Qin, Fuxun Yu, Xiang Chen}
\affiliation{
	George Mason University, Fairfax, Virginia. $\{$zqin, fyu2, xchen26$\}$@gmu.edu
\vspace{1.5mm}
}

\begin{abstract}
Convolutional Neural Networks (CNNs) are used for a wide range of image-related tasks such as image classification and object detection.
However, a large pre-trained CNN model contains a lot of redundancy considering the task-specific edge applications.
Also, the statically pre-trained model could not efficiently handle the dynamic data in the real-world application. 
The CNN training data and their label are collected in an incremental manner.
To tackle the above two challenges, we proposed~\textit{TeAM}-- a task-adaptive incremental learning framework for CNNs in intelligent edge devices.
Given a pre-trained large model,~\textit{TeAM} can configure it into any specialized model for dedicated edge applications.
The specialized model can be quickly fine-tuned with local data to achieve very high accuracy.
Also, with our global aggregation and incremental learning scheme, the specialized CNN models can be collaboratively aggregated to an enhanced global model with new training data.
\end{abstract}

\renewcommand{\shortauthors}{Z. Qin et al.}
% \keywords{Mobile Application, Filter Pruning, Visualization}

\maketitle
\vspace{-2mm}
\section{\textbf{Introduction}}
\vspace{-1mm}
\label{sec:intr}
With the explosive growth of edge devices and intelligent applications, convolutional neural networks (CNNs) have been widely adopted by edge applications for object detection~\cite{Girsh:2014:CVPR}, and image classification~\cite{Kriz:2012:NIPS}.
For current CNN utilization, the models are generally trained with vast classification targets, which are supposed to have sufficient generality to support various intelligent applications.
However, only partial classes are practically needed in edge application scenarios due to individual users’ preference and application specificity, causing unnecessary computation consumption.
Also, the pre-trained model is inability to learn new classification targets since the training process is static and only done once before it is deployed to practical applications.

\begin{figure}[t]
	\begin{center}
	\includegraphics[width=3.3in]{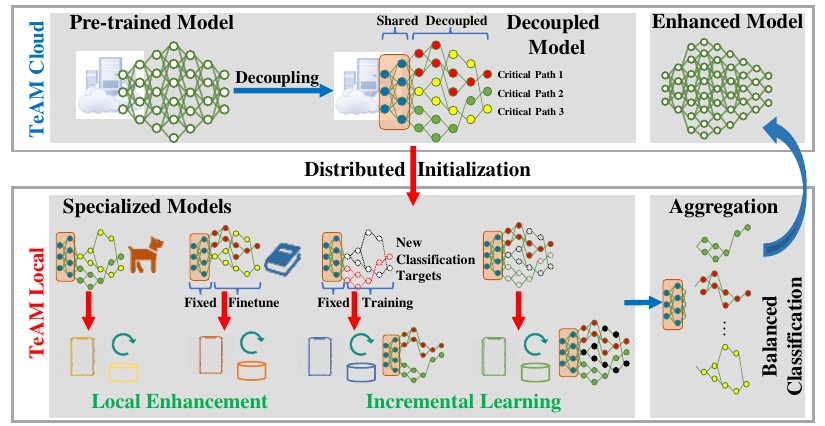}
	\vspace{-2mm}
	\caption{Overview of the Task-Adaptive Incremental Learning Framework.}
	\label{fig:overview}
	\vspace{-4mm}
	\end{center}
\end{figure}

To tackle these challenges, we proposed~\textit{TeAM}-- a task-adaptive incremental learning framework for CNNs in intelligent edge devices.
As shown in Fig.~\ref{fig:overview}, in the cloud, the pre-trained model is decoupled into shared layers and decoupled layers by pruning the unneeded filters and connections.
In the decoupled layers, the CNN are decoupled into individual ``critical paths'', which can be only activated by specific classification targets.
Based on local application specific, specialized models that consist of the shared layers and dedicated critical paths are distributed to local devices for local learning including two learning scenarios: 
(1) \textit{Local Enhancement.} 
When the local devices have the same classification targets as the specialized model, the models are quickly fine-tuned to increase local model performance by using a limited amount of training data and computation resource.
(2) \textit{Incremental Learning.}
To handle the ever-changing environment in local devices, our decoupled models can be incrementally trained with the new classification targets.
The shared layers act as a basic feature extractor for any new local data, and the critical paths can be trained from scratch for additional new classification targets.

Experiments shows that, our ~\textit{TeAM} can achieve 1.7\% and 1.2\% accuracy improvement than the cloud pre-trained AlexNet and VGG model on the imbalanced CIFAR-10 datasets, respectively. 
Meanwhile, the communication traffic of~\textit{TeAM} is significantly lower than federated learning~\cite{mcmahan2016FL}.

% \newpage
\vspace{-3mm}
\section{\textbf{Task-adaptive Local Model Enhancement}}
% \vspace{-1mm}

\vspace{-2mm} 
\subsection{\textbf{Task-Adaptive Model Customization}}
\vspace{-1mm}

\begin{figure}[b]
\vspace{-2mm}
\begin{center}
	\includegraphics[width=3.3in]{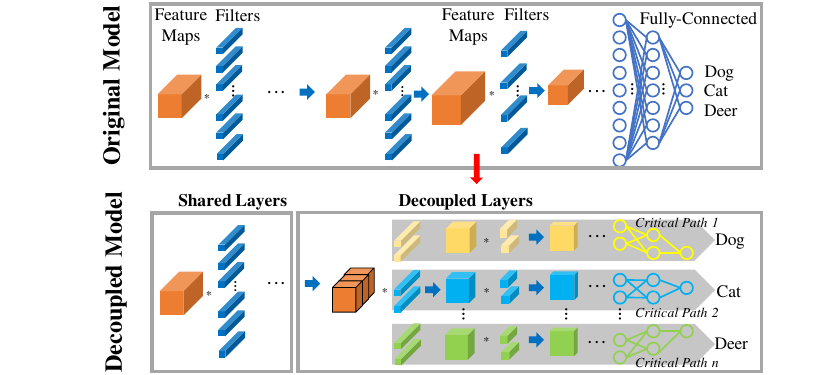}
	\vspace{-4mm}
	\caption{Overview of CNN Decoupling.}
	\label{fig:cp}
	% \vspace{-6mm}
\end{center}
\end{figure}

Our framework is designed to customize a pre-trained CNN model into task-specific models according to user’s preference and application specificity. 
In the~\textit{TeAM} cloud, we leverage the model decoupling technique~\cite{yu2018distilling} to customize the pretrained CNNs.
As shown in Fig.~\ref{fig:cp}, the model decoupling changes an original CNN model’s structure into two parts: shared layers, and decoupled layers: 
(1) The shared layers are kept as the original model structure without decoupling since these layers mainly extract basic features that are generally utilized by all classes. 
(2) In the decoupled layers, the original convolutional layer was decoupled into n independent ``critical paths" across all these layers by removing the filter connections between individual paths.
Every filter in the critical paths only convolves with intra-path feature maps instead of all feature maps produced by the previous layer. 
Based on local application specific, a specialized model that consists of the shared layers with dedicated critical paths is deployed to local devices for local learning.
\vspace{-3mm}
\subsection{\textbf{Local Distributed Model Enhancement}}
\vspace{-1mm}

% \begin{table}[t]
% \vspace{-4mm}
% \centering
% \fontsize{8}{12}\selectfont
% \setlength{\tabcolsep}{3pt}
% \caption{Different Local training Configuration for 10 Classes }
% \vspace{-3mm}
% \label{tab:tune}
% \begin{tabular}{c|c|c|c|c|c|c|c|c|c}
% \whline
% \multicolumn{2}{c|}{Configuration} 		 		& 0    & 100    & 300    & 500  & 700 & 1k   & 8k   & 10k   \\ \hline
% \multicolumn{2}{c|}{Tune FC Layer}       		& 81.8 & 82.3   & 83.6   & 82.9 & 83.1& 82.1 & 88.8 & 90.1 \\ \hline
% \multicolumn{2}{c|}{\textbf{Tune CP FC Layers}} & 81.8 & \textbf{82.5}   & \textbf{83.7}   & \textbf{83.1} & \textbf{83.6} & \textbf{83.8} & \textbf{89.8} & \textbf{91.3} \\ \hline
% \multicolumn{2}{c|}{Tune ALL Layers}            & 81.8 & 82.0   & 82.7   & 82.5 & 81.7 & 82.1 & 87.2 & 89.4 \\ \whline
% \end{tabular}
% \end{table}

\begin{figure}[t]
\vspace{-2mm}
\begin{center}
	\includegraphics[width=3.3in]{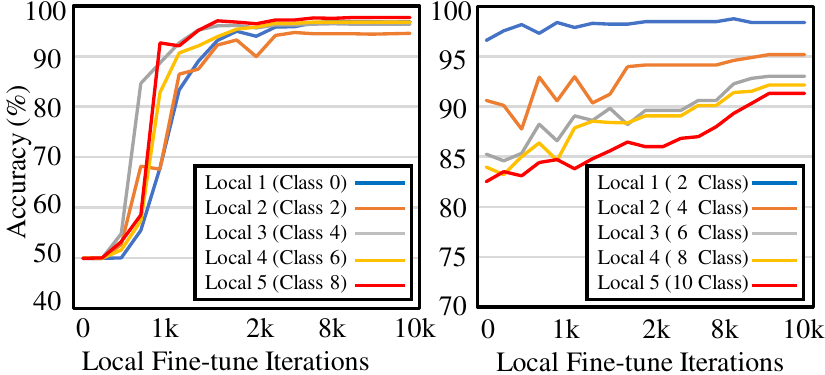}
	\vspace{-2mm}
	\caption{Local Distributed Model Enhancement.}
	\label{fig:tune}
	\vspace{-6mm}
\end{center}
\end{figure}

After the specialized model was deployed into the local devices, the critical paths in the decoupled layers can be quickly fine-tuned with the local data.
In the fine-tuning process, the weights of the shared layers of the specialized model are frozen.
This is because the shared layers capture basic features like edges, lines, and colors, which are not specific to a particular classification target.
% Table~\ref{tab:tune} shows the fine results of a 10 classes specialized model under different configurations.
% We fine-tune the fully-connected layer (Tune FC layer), the decouple layers (Tune CP and FC layer), and all layers (Tune ALL Layers) respectively. 
% As shown in the Table~\ref{tab:tune}, freezing the shared layers while fine-tuning the decouple layers can achieve the best accuracy. 
Fig.~\ref{fig:tune} shows the local learning process when each local device are deployed with a specialized model with a different number of critical paths.
We can see that:
(1) The fine-tuning process can quickly recover specialized model’s accuracy. 
(2) When the local device decoupled with a single critical to perform one-vs-all classification, the model can achieve up 98\% classification accuracy.
(3) With more critical paths in the specialized model, the accuracy will close to the pre-trained model.
In the next section, we will introduce distributed training with incremental learning based on task-adaptive distributed initialization.

\vspace{-3mm}
\section{\textbf{Cloud Model Aggregation with Incremental Learning}}
\vspace{-2mm}
\subsection{\textbf{Global Aggregation with Critical \\Path Average}}
\vspace{-1mm}

\begin{figure}[t]
\vspace{-2mm}
\begin{center}
	\includegraphics[width=3.3in]{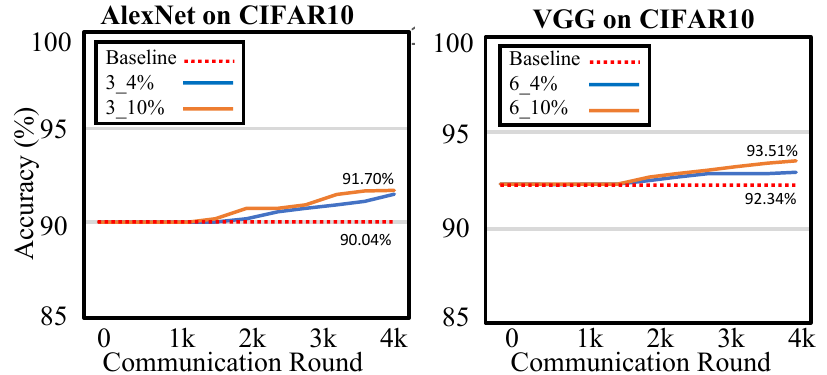}
	\vspace{-2mm}
	\caption{Global Aggregation Evaluation.}
	\label{fig:aggregation}
	\vspace{-6mm}
\end{center}
\end{figure}

In addition to the local enhancement, our proposed framework can also collaboratively update a global shared model. 
First, the~\textit{TeAM} cloud needs the initialization of specialized model by model decoupling. 
Then, the~\textit{TeAM} cloud starts a new training round $t$ and sends the specialized model to the local devices. 
Next, each local devices trains the specialized model with the mini-batch SGD for E local epochs on the local data and calculates the updates of the critical path $w_{cp}(t)$ in parallel.
The local epoch E affects only the time spent on training per local device and does not increase additional communication overhead. 
Finally, the~\textit{TeAM} collects the updates of all local devices, aggregates the updates with the average of all corresponding critical paths, then updates the global model to $w_{cp}(t+1)$ and ends this round. 
$w_{cp}(t+1)$ is the start model for the next round.
Fig.~\ref{fig:aggregation} shows the global aggregation evaluation for the AlexNet and VGG on the CIFAR10 dataset. 
3\_4\%means the last 3 convolutional layers are decoupled with 4\% of filters in each critical path. 
We can achieve higher classification accuracy than the pre-trained model.

\vspace{-2mm}
\subsection{\textbf{Incremental Learning with New Classification Targets}}
\vspace{-1mm}

To handle the ever-changing environment in local devices and achieve an ideal accuracy, the CNN models should be incrementally trained with the new local data.
Traditionally, we need to change the model structure and utilize all the data to retrain the whole model.
However, with the model decoupling, we can easily add additional critical path on the existing specialized models, which can be trained to handle the new data.

For incremental learning with n old class and m new classes, we learn a new model to perform classification on n + m classes, by using the shared layers from an old model that classifies the old n classes. 
Specifically, we can deploy the initialized shared layers with random initialized critical paths to local device. 
However, during the training procedure, it is critical to reconcile the learning rate of the new critical paths and old paths. 
We propose an adaptive learning rate scheme to different critical paths, which can significantly improve the global model accuracy. 
We empirically set the learning rate of the new critical path is 10 times to the old paths.
7 class targets from the CIFAR10 are gradually added on the baseline model trained with a subset of CIFAR100 (e.g. \textit{e.g.} 10, 20, 30).  
Although there is accuracy drop after additional class targets are added on the baseline model, the accuracy of the incremented class targets is comparable to the baseline model.

% For the AlexNet model, the global model can achieve 91.7\% classification accuracy under 4k communication round. 
% For the VGG model, the global model can achieve 93.51\% classification accuracy under 500 communication round. 
% In this way, our local devices can improve its accuracy on dynamic data in real local mobile environments with the reduced computation load.

% \vspace{-3mm}

\vspace{-3mm}
\section{Conclusion}
\vspace{-2mm}
In this paper, we proposed a novel task-adaptive incremental learning framework for deep learning based edge applications. 
We first customize the pre-trained large model into specialized models based local classification preference. 
Then, two learning scheme are proposed to improve local model performance and collaborated update the global model.

\vspace{-2mm}

\let\oldbibliography\thebibliography
\renewcommand{\thebibliography}[1]{%
  \oldbibliography{#1}%
  \setlength{\itemsep}{0pt}%
}
\footnotesize
\bibliographystyle{IEEEtran}
\bibliography{SEC_team}

\end{document}